# Computer simulation of effective viscosity of fluid-proppant mixture used in hydraulic fracturing

Vitaly A. Kuzkin, Anton M. Krivtsov, and Aleksandr M. Linkov

*Rzeszow University of Technology,*
*Saint Petersburg State Polytechnical University,*
*Institute for Problems of Mechanical Engineering, Russian Academy of Sciences*
E-mail: kuzkinva@gmail.com

**Abstract**

The paper presents results of numerical experiments performed to evaluate the effective viscosity of a fluid-proppant mixture, used in hydraulic fracturing. The results, obtained by two complimenting methods (the particle dynamics and the smoothed particle hydrodynamics), coincide to the accuracy of standard deviation. They provide an analytical equation for the dependence of effective viscosity on the proppant concentration, needed for numerical simulation of the hydraulic fracture propagation.

## 1. INTRODUCTION

Hydraulic fracturing technology is used for stimulation of oil and gas production [1]. The pioneering results on mathematical modeling of hydraulic fractures were given by Khristianovich and Zheltov [2, 3]. Further development of the analytical and numerical methods has been reviewed in many papers (e.g. [4-11]). In all the studies and computer codes, used for simulation of the final stage of fracturing, the mixture of a fluid and proppant is modeled as a single fluid with the density and viscosity depending on proppant concentration (see, e.g. [6]). While there is no problem with defining the efficient density, prescribing viscous properties presents a problem uneasy to solve. As noted in the paper [6], "Regarding the viscosity of the slurry, this is actually one of the most difficult (and critical) aspects of the modeling. Proper formulation of the momentum equation for the problem of a suspension of solid particles yields terms that are related to the interaction between particles and between particles and the fluid. Accounting for these effects in detail is challenging and most models that attempt to describe these interactions are still awaiting experimental verification".

A variety of models for dependence of effective viscosity $\mu_s$ on particle concentration $c$ has been suggested. The asymptotical behavior of the viscosity at small concentrations is described by Einstein formula [12]:

$$\mu_s = \mu_s(0)(1 + Ac), \tag{1}$$

where $A = 5/2$ in 3D, while $A = 2$ in 2D problems [13]. Equation (1) does not take into account hydrodynamic interactions between proppant particles and therefore it is not applicable at high concentrations. More complicated models have been proposed by Mooney [14], Maron and Pierce [15], Krieger and Dougherty [16]:

$$\mu_s^M = \mu_s(0) \exp \frac{Ac}{(1 - c/c_*)}, \qquad (2)$$

$$\mu_s^{MP} = \frac{\mu_s(0)}{(1 - c/c_*)^2}, \qquad (3)$$

$$\mu_s^{KD} = \frac{\mu_s(0)}{(1 - c/c_*)^{Ac_*}}, \qquad (4)$$

where $c_*$ is a critical concentration, commonly used as a fitting parameter, $A$ is the Einstein coefficient in (1). The model (4) is used for simulation of proppant transport in hydraulic fractures, for example, in the paper [6]. Note that formulas (2)-(4) suggest qualitatively different dependencies of viscosity on proppant concentration. The choice between the models is not straightforward. According to the review [17], the majority of the models are derived either analytically or by fitting experimental data. Evidently each of the approaches has its limitations. Analytical models usually incorporate strong assumptions of limited applicability. The challenges of experimental techniques are described in the paper [18].

Consequently, computer simulations may play an important role as an additional tool for the investigation. The solution of Navier-Stockes equations for the suspension using conventional methods of computational fluid dynamics is extremely time-consuming. Therefore many alternative techniques, such as Stokesian dynamics [19], dissipative particle dynamics [20], smoothed particle hydrodynamics [21], molecular dynamics [22], lattice Boltzmann [23], etc., are used in literature for simulation of suspensions.

In this paper, the particle dynamics (PD) [24, 25] and smoothed particle hydrodynamics (SPH) [26-28] are used. These methods mutually complement each other. The PD is simple and it contains a small number of parameters. However in the framework of the PD the viscosity cannot be specified explicitly. On the other hand, in the SPH, viscosity is a parameter of the model. At the same time the motion of smoothed particles in some cases is artificial [28]. Therefore the joint use of these methods may serve for verifying the results and for better understanding of the suspension behavior.

The study of numerical simulation of the Poiseuille flow of a suspension in a narrow channel by the two complimenting methods has been initiated in the paper [29]. Meanwhile, the results of [29] referred to the proppant concentration not exceeding 0.3. It has appeared that considering higher concentrations required much greater computational effort because of the need to consider systems with notably greater number of degrees of freedom (DOF). In this work paper, we increased the number of DOF to the level providing reliable results up to the concentration 0.6, which is close to the ultimate concentration of randomly packed particles. The numerical results obtained serve us to compare the analytical models (2)-

(4) and to choose that model, which complies with the results of numerical experiments. We conclude that the Maron-Pierce equation (3) with $c_*^{MP} = 0.77$ is the best-fit one.

## 2. STATEMENT OF THE PROBLEM. SIMULATION TECHNIQUES

In this section we briefly summarize the statement of the problem. We study the flow of a Newtonian fluid containing proppant particles in a channel of constant width. The channel is simulated by a square computational domain with periodic boundary conditions [25] in the direction of the flow and rigid walls in the orthogonal direction. The rigid walls are simulated by using two rows of fixed fluid particles. The flow is driven by the constant body force acting along the flow. It is shown in paper [29] that this statement is equivalent to the flow under constant pressure gradient. This serves us to simplify modeling by the both methods, because simulation of the body force is notably simpler than prescribing pressure gradient.

The fluid particles initially form a perfect square lattice with nearest neighbor distances equal to $a_0$. Proppant particles are either distributed randomly with uniform spatial distribution (for volumetric concentration $c \leq 0.4$) or form square lattice (for $c > 0.4$). The volumetric concentration $c$ is defined as

$$c = \frac{\pi R^2 N_p}{L^2},$$

where $L$ is the size of the computational domain; $N_p$ is the total number of proppant particles; $R$ is the radius of a proppant particle.

The initial velocities correspond to the Poiseuille flow of a single Newtonian fluid. In the case of the PD simulations, fluid particles have additional random velocities. The presence of proppant changes the rheological properties of the suspension, specifically it increases the effective viscosity. This leads to decaying the initial parabolic profile of the in-plane particle velocity until the steady-state regime is reached. The velocity $v_{av}$ of the center of mass of all particles inside the computational domain is calculated during the PD and SPH simulations. In a steady-state regime $v_{av}$ is identical to average profile velocity. The effective viscosity and the effective density of the suspension are defined by equations [26]:

$$\mu_s = \frac{\rho_s g L^2}{12 v_{av}}, \quad \rho_s = \frac{m N_f + m_p N_p}{L^2}, \qquad (5)$$

where $g$ is the body force; $m$, $N_f$ are the mass and the total number of fluid particles; $m_p$ is the mass of proppant particle. The first of equations (5) corresponds to the Poiseuille flow of a Newtonian fluid under the action of the body force $g$. In computer simulations the body force was renormalized so that $\rho_s g$ does not depend on the proppant concentration. In this case the only parameter in the first of equations (5), depending on the proppant concentration, is the average velocity $v_{av}$. In view of (5), it defines the effective viscosity.

The key-features of the simulation techniques are summarized below.

## 2.1. PARTICLE DYNAMICS

The first method used in the present paper is the particle dynamics [24, 25]. In the framework of the PD method, Newtonian equations of motion for interacting particles, representing both the fluid and proppant, are solved numerically. In the present paper symplectic leap-frog [30] integration scheme is used. The particles interact via the spline potential [25]. The force, acting between fluid particles $i$ and $j$, is calculated as follows:

$$\mathbf{F}_{ij} = \frac{fk(r_{ij})}{a}\left(\left(\frac{a}{r_{ij}}\right)^8 - \left(\frac{a}{r_{ij}}\right)^{14}\right)\mathbf{r}_{ij}, \quad \mathbf{r}_{ij} = \mathbf{r}_j - \mathbf{r}_i,$$

$$k(r) = \begin{cases} 1, & r < b, \\ \left(1 - \left(\frac{r^2 - b^2}{a_{cut}^2 - b^2}\right)^2\right)^2, & b \leq r < a_{cut}, \\ 0, & r \geq a_{cut}, \end{cases} \quad (6)$$

where $a_{cut}$ is a cut-off radius, $a$ is an equilibrium distance between particles, $f$ is a force constant, $b = (13/7)^{1/6} a$. Following the approach, proposed in the paper [29], we represent every proppant particle as a set of rigidly connected smaller particles shown in figure 1. Thus each proppant particle is a rigid body with two translational and one rotational degree of freedom. The distance between the nearest particles for the outer circle is equal to the equilibrium distance $a$ between fluid particles. The distance between inner and outer circles is also equal to $a$. These particles interact with fluid particles via the forces defined by equation (6).

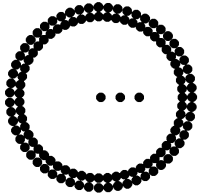

**Fig. 1.** Representation of a proppant particle as a set of rigidly connected smaller particles. Three particles in the middle are used for visualization of particle orientation.

Note that the interactions with fluid cause both translation and rotation of proppant particles. Thus the equations of motion of a proppant particle $i$ have the form

$$m_p \dot{\mathbf{v}}_i = \sum_{k \in \Lambda_i, j \notin \Lambda_i} \mathbf{F}_{kj} + m_p \mathbf{g}, \qquad \Theta_p \ddot{\boldsymbol{\varphi}}_i = \sum_{k \in \Lambda_i, j \notin \Lambda_i} (\mathbf{r}_k - \mathbf{r}_i) \times \mathbf{F}_{kj}, \quad (7)$$

where $m_p$, $\Theta_p$ are mass and moment of inertia with respect to the center of mass of a proppant particle; $\boldsymbol{\varphi}_i$ is the angle describing the orientation for the $i$-th proppant particle; $\Lambda_i$ is the set of indices for the particles representing the proppant particle $i$; $\mathbf{g}$ is the body force driving the flow. In paper [29], it has been shown that this approach is computationally more efficient than the straightforward approach, when spheres of different size are used for representation of the proppant and fluid particles.

In the PD method properties of the fluid depend on the number of particles being used. Consider two discrete systems, marked by the subscripts 0 and 1, with different number of particles $N_0$ and $N_1$ corresponding to the same square specimen of the fluid. The relation between parameters of these discrete systems is found as follows. Assuming that the geometrical size of the system, fluid density, and the sound speed $v_s = \sqrt{6fa/m}$ do not depend on the number of particles, we have:

$$a_1 = \sqrt{\frac{N_0}{N_1}} a_0, \quad m_1 = \frac{N_0}{N_1} m_0, \quad f_1 = \sqrt{\frac{N_0}{N_1}} f_0, \tag{8}$$

where $a_i, m_i, f_i$ ($i = 0, 1$) are equilibrium distance, particle mass, and force constant for the system $i$. In the paper [29], it has been stated that when having the relations (8) satisfied, the viscosities of the discrete systems are related as:

$$\mu_1 = \sqrt{\frac{N_0}{N_1}} \mu_0.$$

Thus the viscosity of the system decreases with increasing resolution (number of particles). However one can avoid the analogous dependence for the Reynolds number. It is straightforward to show that the body force should have the form (see the paper [29] for more details):

$$g_1 = \frac{N_0}{N_1} g_0.$$

The procedure described allows us to avoid the dependence of the main parameters of the problem (size, density, sound speed, and Reynolds number) on the total number of particles in the system.

The heat generated by shear flow is removed from the system by using the Berendsen thermostat [31]. The thermostat is applied to a narrow fluid strip (of $5a$ width) near the left boundary of the computational domain. Then the heated fluid, leaving the domain through its right side, is cooled down by the thermostat after crossing the left side of the domain.

## 2.2. SMOOTHED PARTICLE HYDRODYNAMICS

The second method used in the present paper for simulation of the proppant transport is the smoothed particle hydrodynamics [26-28]. Similarly to the PD, in the SPH, a fluid is represented by a set of interacting particles. The motion of the smoothed particles is governed by the equations:

$$\dot{\mathbf{v}}_i = -\sum_j m \left( \frac{p_i}{\rho_i^2} + \frac{p_j}{\rho_j^2} + S_{ij} \right) w'(r_{ij}) \mathbf{e}_{ij} + \mathbf{g}, \quad \rho_i = \sum_j m w(r_{ij}),$$

where $p_i, \rho_i$ are, respectively, the pressure and the density at the point, where the particle $i$ is located; $S_{ij}$ is a viscous term; $w$ is a weighting function. The weighting function $w(r)$ has a compact support, vanishing for $r \geq a_{\text{cut}}$, where $a_{\text{cut}}$ is a smoothing length similar to the cut-off radius used in the particle dynamics method. We employ the Lucy weighting function [26]:

$$w(r) = \frac{5}{\pi a_{cut}^2}\left(1 + 3\frac{r}{a_{cut}}\right)\left(1 - \frac{r}{a_{cut}}\right)^3, \quad r \in [0; a_{cut}].$$

and the constitutive relations by Monaghan [27] for the pressure and the viscous term:

$$p_i = B\left(\left(\frac{\rho_i}{\rho_0}\right)^\gamma - 1\right), \quad S_{ij} = -\frac{16\mu_i\mu_j}{\rho_i\rho_j(\mu_i+\mu_j)}\frac{\mathbf{v}_{ij}\cdot\mathbf{r}_{ij}}{r_{ij}^2+\varepsilon a^2}, \quad \mu_i = \frac{1}{8}\alpha a_{cut}v_s\rho_i, \quad \mathbf{v}_{ij} = \mathbf{v}_j - \mathbf{v}_i. \quad (9)$$

Herein, $\rho_0$ is the equilibrium fluid density; $B$, $\alpha$, $\gamma$, $\varepsilon$ are parameters of the model; $a$ is a characteristic size of the particle; $v_s$ is the speed of sound. In the paper [27], it has been shown that the equation of state for the pressure in the form (9) guarantees low compressibility of the fluid. In contrast to the PD, where viscosity arises naturally as a result of stochastic motion, in the SPH, the viscosity is introduced explicitly as the key parameter of the model. Additionally the following purely repulsive core potential is used for preventing the formation of artificial structures in the fluid [28]:

$$\Phi_{core}(r) = \frac{fa_{cut}}{4}\left(1 - \frac{r^2}{a_{cut}^2}\right)^4,$$

where $a_{core}$ is a cut-off radius for the core potential. Interactions between proppant particles, as well as between proppant and fluid particles, are described by (6). For proppant-proppant interactions the forces are truncated at $r = a$, hence the interactions are purely repulsive. The motion of proppant particles is governed by equations (7).

Consider the creation of an initial configuration. The fluid described by the equation of state (9) is nearly incompressible. Therefore the computational domain should be completely filled by the particles. Otherwise the system would contain artificial voids, similar to gas bubbles.

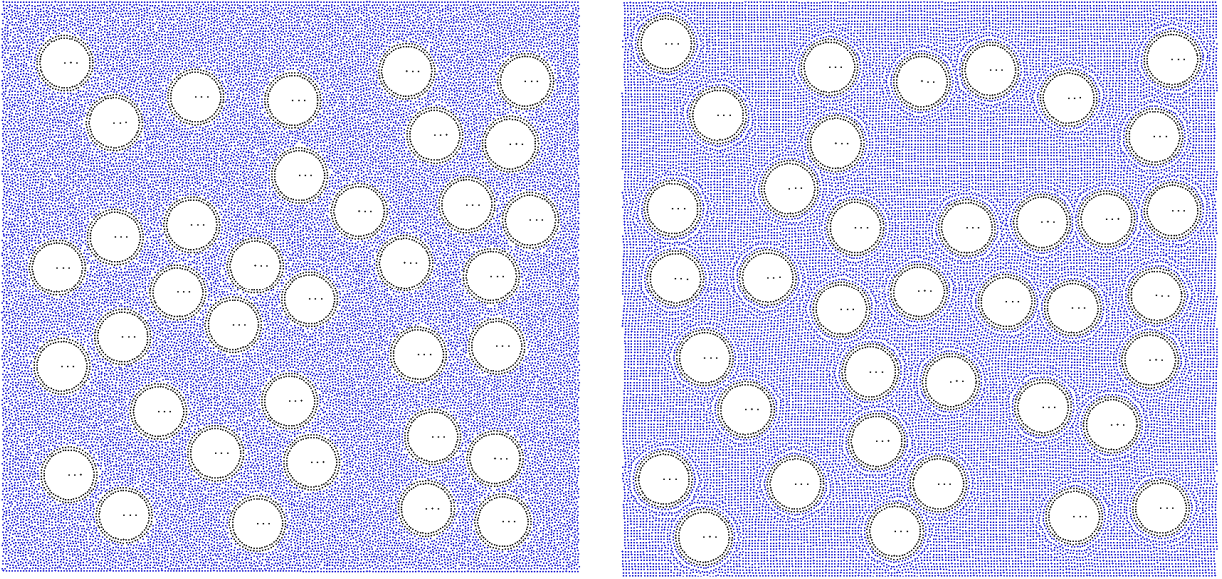

**Fig. 2.** The distribution of fluid and proppant particles for c = 0.3: particle dynamics (left) and smoothed particle hydrodynamics (right).

In the present paper, the following equilibration procedure is used. Proppant particles are set as described above. The remaining space is filled by fluid particles forming perfect square lattice. Evidently in this case some voids are formed around proppant particles. In order to remove the voids, the system is compressed by multiplying equilibrium density $\rho_0$ by 0.8. After that, the density is slowly increased until the pressure in the system reaches the value of $0.01K$, where $K$ is the bulk modulus of the fluid. In the course of this procedure the fluid and proppant particles move in accordance with the equations of motion described above.

The resulting distribution of particles after equilibration is shown in figure 2 (right). One can see that the computational domain is completely filled by the particles and no significant artifacts are present. After the equilibration, the particle velocities are set in accordance with parabolic velocity profile corresponding to the Poiseuille flow.

### 3. CHOICE OF MODEL PARAMETERS

The main dimensionless parameters influencing relative viscosity of the suspension are the Reynolds number Re, the ratio of the proppant diameter to the channel width, and the ratio of the proppant density to the fluid density. We calibrate the model so that the given parameters correspond to the flow of the proppant-fluid mixture in hydraulic fracture. The most widespread size of proppant particles is 20/40 mesh (0.4-0.8 mm). The opening of the crack is of 10 mm order. To estimate the typical Reynolds number for hydraulic fracturing we assume that the fracturing fluid is water with the density and dynamic viscosity at normal conditions being 1000 kg/m$^3$ and 0.0009 Pa·s, respectively. The characteristic velocity of the flow is 0.01 m/s. Therefore the typical Reynolds number is of order of 1. Note that this number should not be fitted exactly in computer simulations. The only requirement for the simulation is that the flow is laminar. Therefore an order higher Reynolds numbers may be used to speed up the simulations. We employ the following values of parameters used in the framework of the both methods:

$$\frac{R}{L} = \frac{1}{20}, \quad \frac{\rho_p}{\rho_f} = 2, \quad \text{Re} \approx 30, \quad v_s = 2.45, \quad m = 0.09, \quad a = 0.3, \quad f = 0.3,$$

where $L$ is the size of the computational domain; $R$ is the radius of a proppant particle; $\rho_f$, $\rho_p$ are the fluid and proppant densities, respectively; $v_s$ is the sound speed.

The specific values of the parameters used in the PD simulations are:

$$\frac{T}{fa} = 0.125, \quad \frac{a_{cut}}{a} = 2.1, \quad \frac{\Delta t}{t_*} = 0.01, \quad \frac{a_0}{a} = 0.98, \quad \frac{v_0}{v_s} = 0.14, \quad \frac{mg}{f} = 2.58 \cdot 10^{-5},$$

where $T$ temperature (kinetic energy); $\Delta t$ is the time step; $t_* = 2\pi\sqrt{ma/6f}$; $a_0$ is the initial distance between fluid particles; $v_0$ is the amplitude of initial random velocities of fluid particles.

In the SPH simulations, the parameters are:

$$\frac{B}{\rho_0 v_s^2} = \frac{1}{7}, \quad \alpha = 0.5, \quad \frac{\beta}{a_{cut} v_s} = 0.23, \quad \gamma = 7, \quad \varepsilon = 0.001,$$

$$\frac{a_{cut}}{a} = 2.5, \quad \frac{\Delta t}{t_*} = 0.02, \quad \frac{a_0}{a} = 1, \quad \frac{v_0}{v_s} = 0.01, \quad \frac{mg}{f} = 6 \cdot 10^{-5}.$$

The simulations have been carried out at the Department "Theoretical Mechanics" of Saint Petersburg State Polytechnical University by using the supercomputer KS-EVM-1TF. It has 144 cores and peak performance of 1 Tflops. The approximate number of particles and the number of time steps used in simulations are:

$$N = 4 \cdot 10^4, \quad s_{max} = 2 \cdot 10^6.$$

The choice of the number $N$ of particles and the number $s_{max}$ of time steps strongly depends on computational facilities available. For each proppant concentration five problems with different initial conditions are solved by two methods. One core per a problem is used. The numerical results presented below correspond to 130 simulations and approximately two weeks work of the supercomputer.

## 4. DISCUSSION OF NUMERICAL RESULTS ON EFFECTIVE VISCOSITY OF FLUID-PROPPANT MIXTURE

The effective viscosity of the suspension, calculated for different proppant concentrations by using the particle dynamics and the smoothed particle hydrodynamics, is shown in figure 3. The results are normalized by the viscosity $\mu_s(0)$ of a pure fluid. Every point on the plot is the mean of five simulations with different initial proppant distributions. The bars on the plot show the dispersion of the results (average value plus/minus standard deviation). The solid line corresponds to the Einstein formula (1) in 2D case ($A = 2$). It can be seen that the difference between the results of the PD and the SPH simulations is of order of dispersion of the PD results. For proppant concentration higher than 0.2, the obtained values of the suspension viscosity are higher than the value predicted by the Einstein formula (1). Therefore at these concentrations the hydrodynamic interactions between proppant particles, neglected in the Einstein derivation, are significant. In this case, the non-linear equations (2)-(4) are to be used.

The critical concentration, entering (2)-(4), has been used as a fitting parameter. For it, the following values were obtained by applying the least square method: $c_*^M = 0.99$, $c_*^{MP} = 0.77$, $c_*^{KD} = 0.67$, for equations (2), (3) and (4), respectively. The corresponding curves are also shown in figure 3. It can be seen that equations (2) and (3) give better approximation of the numerical results than the equation (4). Note now that the critical concentration $c_*^M = 0.99$, fitting the approximation (2), is unrealistic, since the highest concentration in 2D, corresponding to a triangular lattice, is $\pi/(2\sqrt{3}) \approx 0.91$. In contrast, the critical concentration 0.77, obtained for the approximation (3), is quite close to the concentration corresponding to random close packing in two dimensions $c_{RCP} = 0.82 \pm 0.02$ [32].

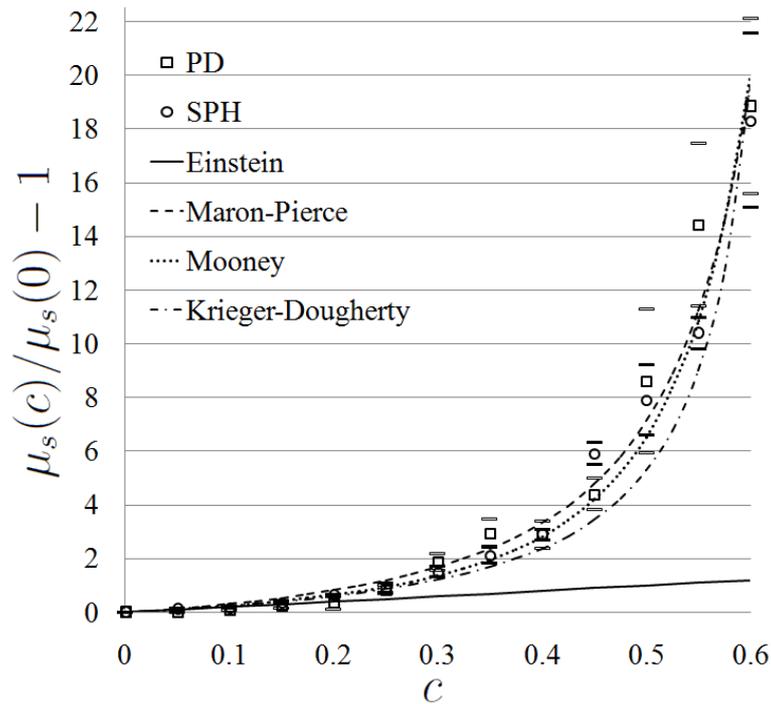

**Fig. 3.** The dependence of relative viscosity on proppant concentration obtained using the PD, the SPH, and formulae (1)-(4).

This implies that the Maron-Pierce equation (3) provides the best fit for simulation results, obtained by two complimenting methods, in the entire range of the proppant concentration. This result is in good agreement with the experimental study [18], where it has been shown that the Maron-Pierce model accurately predicts effective viscosity of the suspension. Thus we conclude that equation (3) may be recommended for simulation of proppant transport in hydraulic fractures.

## ACKNOWLEDGEMENT

The support of the European Research Agency (FP7-PEOPLE-2009-IAPP Marie Curie IAPP transfer of knowledge program (Project Reference No. 251475)) is gratefully acknowledged.